# Combinatorial Testing for Deep Learning Systems

Lei Ma[1,2], Fuyuan Zhang[2], Minhui Xue[3], Bo Li[4],
Yang Liu[2], Jianjun Zhao[5], and Yadong Wang[1]

[1] Harbin Institute of Technology, China
[2] Nanyang Technological University, Singapore
[3] New York University Shanghai, China
[4] University of Illinois at Urbana–Champaign, USA
[5] Kyushu University, Japan
Contact author: `malei@hit.edu.cn`

**Abstract.** Deep learning (DL) has achieved remarkable progress over the past decade and been widely applied to many safety-critical applications. However, the robustness of DL systems recently receives great concerns, such as adversarial examples against computer vision systems, which could potentially result in severe consequences. Adopting testing techniques could help to evaluate the robustness of a DL system and therefore detect vulnerabilities at an early stage. The main challenge of testing such systems is that its runtime state space is too large: if we view each neuron as a runtime state for DL, then a DL system often contains massive states, rendering testing each state almost impossible. For traditional software, combinatorial testing (CT) is an effective testing technique to reduce the testing space while obtaining relatively high defect detection abilities. In this paper, we perform an exploratory study of CT on DL systems. We adapt the concept in CT and propose a set of coverage criteria for DL systems, as well as a CT coverage guided test generation technique. Our evaluation demonstrates that CT provides a promising avenue for testing DL systems. We further pose several open questions and interesting directions for combinatorial testing of DL systems.

## 1 Introduction

Deep learning (DL) systems have been widely applied in various applications due to their high accuracy, such as computer vision [35], natural language processing [39], autonomous driving [5], and automated medical diagnosis [33]. However, recently DL systems have been shown to be vulnerable against different attacks, such as adversarial examples in computer vision and audio systems. Given that more and more safety-sensitive applications start to adopt DL, deploying DL without thorough testing to safety-critical applications can lead to severe consequences, such as possible accidents in autonomous driving [12]. DL systems are notoriously difficult to debug because of the complex gradient computation (such as back-propagation). This raises the question of how to efficiently verify or test DL effectively. The attempt to verify DL is exceptionally challenging since the size of runtime states of deep neural networks (DNNs) is often huge if we view each neuron output as a state, making existing formal verification techniques intractable. Traditional software testing techniques are considered to be able to provide system defect detection at an early stage, which is promising to enhance the system robustness. Testing is also a widely-used technique in hunting bugs and evaluating the quality of traditional software systems. Research on designing test coverage and test case generation for DL is emerging [30,34]. However, these work mainly rely on probing the accuracy on test data which are randomly drawn from manually labeled datasets and *ad hoc* simulations [38], and some of them focus on small-scale neural networks [34]. Besides, neuron coverage criteria have been proposed recently and have empirically demonstrated to be effective for guiding test case generation [25], while the synergy effects of neurons within and across layers (e.g., massive runtime states) still remain unknown, which is important to the robustness evaluation of DL systems. Therefore, we aim to efficiently test large-scale DL systems based on reasonable size of representative test inputs.

Combinatorial testing (CT), which is used to sample test inputs and configurations from the original space, can well trade-off the defect detection ability and the size of test input. CT can detect faults that are otherwise undetectable effectively, and has been successfully applied to different configurable software systems [2,23,28], as most faults are caused by interactions involving only a few parameters [20]. For example, A $t$-way combinatorial test set is designed to cover all the interactions involving no more than $t$ input parameters. When input parameters are properly modeled, a $t$-way test set enables to expose faults efficiently [9].

In this paper, we propose to apply CT to testing massive runtime states (neuron outputs) of DL systems. However, covering the output space of a layer of neurons in DL is challenging because (1) exhaustive testing is infeasible to enumerate all output values of a neuron; (2) the number of output values of a neuron is also infinite. To sidestep, we discretize the space of the output values into intervals such that each interval is covered. This way, the number of combinations of intervals in their output space can be finite but still increases exponentially with the number of neurons. We, therefore, adopt CT to sample the neuron interactions inside different layers and reduce the number of test inputs that has to be executed.

This paper performs an exploratory study of CT on DL systems. We adapt the concepts in CT and propose a set of CT coverage criteria for DL systems, as well as a CT coverage guided test generation technique. Our evaluation demonstrates that CT is a promising direction for testing DL systems.

The contributions of the paper are summarized as follows:

– We adapt the concept of CT in traditional software testing and propose a set of CT coverage criteria specialized for DL systems.

- We also provide a general CT coverage guided test generation technique for DL systems.
- We perform extensive empirical studies using linear programming based test generation to demonstrate the usefulness of the proposed DL testing techniques. We highlight that CT can provide a promising avenue for testing DL systems.

To the best of our knowledge, this is the first work to explore the feasibility and usefulness of CT on DL systems.

## 2  Related Work

In this section, we will review basic combinatorial testing (CT) techniques, as well as point out the challenges of directly applying CT to DL systems.

### 2.1  Combinatorial Testing

Ensuring the quality and correctness of a system is challenging due to various configurations and input domains for each program. Usually, to test a system demands exhaustive evaluation of all possible configurations and input interactions against their expected outputs. However, such an exhaustive testing effort is infeasible considering the runtime and computation resource limitations. Thus, sampling mechanisms are applied to reduce the dimension of test input, which we call black-box sampling based combinatorial testing (CT). CT has been successfully applied to testing different configurable software systems [2, 23, 28], since most faults are caused by interactions involving only a few parameters [20]. Despite the high testing efficiency, CT is limited in handling various constraints. The ability to handle constraints is crucial for real-world applications since most of the real-world systems are subjected to constraints involving input parameters and configurations. Hence, research on CT [3] has experienced a shift from traditional CT to the constrained CT which breakdowns as meta-heuristic approaches [10, 13, 17, 22], SAT-based approaches [27, 40], and greedy approaches, which is further categorized into one-test-at-a-time (OTAT) [9–11] and in-parameter-order generalized (IPOG) approaches [6, 41, 42].

In summary, CT is usually adopted in practice to address the problem of high cost for test execution. In this paper, we take a step towards exploring whether CT is helpful for testing DL systems.

### 2.2  Deep Learning

Testing [24, 25, 30, 34] and formal verification [14, 18, 31, 32, 36, 37] have been recently applied to deep neural networks. Specifically, for testing DNNs, DeepXplore [30] proposes a white-box differential testing algorithm to systematically generate adversarial examples that cover all neurons in the network. DeepCover [34] proposes the MC/DC test criteria for DNNs. Their test criteria have been evaluated on small-scale neural networks (with only Dense layers, with no more than 400 neurons). DeepGauge [25] proposes a comprehensive set of numerical-based testing criteria (e.g., boundary coverage criteria, major functional coverage critiera), whose usefulness and scalability have been demonstrated on practical-sized DNNs such as ResNet-50 (with near 200 layers and 100,000 neurons). Besides these initial work on proposing testing coverage criteria for DNNs, DeepMutation [24] explores and proposes the first mutation testing technique for DNNs, it provides a general mutation testing framework, including both source-level mutation testing and model-level mutation testing. Sharing the same spirit of mutation testing by injecting minor faults into the DNNs, Breier et.al. [7] recently propose the first laser-based practical fault attacking technique for DNNs.

For verifying DNNs, most recent work $AI^2$ [36] proposes the verification approach on DL systems based on abstract interpretation, and designs specific abstract domains and transformation operators. $AI^2$ is able to handle large neural networks. However, most of the existing testing and formal verification techniques have been evaluated on simple DNN architectures. Enhancing the performance of the existing verification techniques and designing more scalable verification methods towards complex real-world DNNs would be important to improve the robustness of DL systems.

This paper takes the first attempt to study the feasibility and usefulness of CT to testing massive runtime states of Dl systems. We show that generating sufficient test set of reasonable size and test effectiveness can be achieved in parallel by leveraging CT, which has not been examined previously.

## 3   Combinatorial Testing Criteria for DL

We use $N = \{n_1, n_2, \ldots\}$ to represent the set of neurons of a DL system, and $T = \{\mathbf{x}_1, \mathbf{x}_2, \ldots\}$ the set of test inputs. Let $\phi(\mathbf{x}, n)$ be a function that returns the output value of a neuron $n \in N$ given a test input $\mathbf{x} \in T$. For a DNN with $l$ layers, we use $L_i$ to denote the set of neurons on its $i$-th layer ($1 \leq i \leq l$). Our CT testing coverage criteria for DNNs are motivated by the combinatorial coverage metrics in traditional software testing.

We divide the range of the output value of a neuron into two intervals $(-\infty, 0]$ and $(0, \infty)$. A neuron $n$ is activated (resp. deactivated) given $\mathbf{x}$ if $\phi(\mathbf{x}, n) \in (0, \infty)$ (resp. $\phi(\mathbf{x}, n) \in (-\infty, 0]$). We use $A(n_i, \mathbf{x}) \in \{0, 1\}$ to represent the activation status of a neuron given $\mathbf{x}$, where 1 corresponds to *activation* and 0 *deactivation*. To test the functionality of a layer, we first introduce the definition of *neuron-activation configuration*.

**Definition 1 (Neuron-activation configuration).** *For a set of neurons $L_i = \{n_1, n_2, \ldots, n_k\}$, a neuron-activation configuration is a tuple $c = (b_1, b_2, ..., b_k)$, where $b_i = \{0, 1\}$.*

A configuration $c = (b_1, b_2, ..., b_k)$ of $L_i$ is covered by $T$ if there exists a test input $\mathbf{x} \in T$, such that $b_i = A(n_i, \mathbf{x})$ for $1 \leq i \leq k$. We use $FC\_NA(T, L_i)$ to represent that all neuron-activation configurations of $L_i$ are covered by $T$, and $NC\_NA(T, L_i)$ to represent the number of neuron-activation configurations of $L_i$ that are covered by $T$.

In practice, we adapt the concepts of $t$-way combination coverage to reduce the number of necessary testing configurations when testing DL systems.

**Definition 2 ($t$-way combination sparse coverage).** *Given a test set $T$ and a set of neurons $L_i$, we use $\Theta$ to denote the set of all $t$-way combinations of neurons in $L_i$. Then the $t$-way combination sparse coverage represents the percentage of $t$-way neuron combinations in $L_i$, of which all the neuron-activation configurations are covered by $T$.*

$$\text{TWsCov}(T, L_i) = \frac{|\{\theta \in \Theta |\ FC\_NA(T, \theta)\}|}{|\Theta|}$$

*Example 1.* Let $L_i = \{n_1, n_2, n_3, n_4\}$ be a set of neurons in the same layer. Each row $j$ in Table 1 corresponds to the neuron activation status given one test input of $T$, $A(L_i, \mathbf{x}_j)$. There are in total six 2-way combinations of neurons in $L_i$, $\{n_1, n_2\}$, $\{n_1, n_3\}$, $\{n_1, n_4\}$, $\{n_2, n_3\}$, $\{n_2, n_4\}$ and $\{n_3, n_4\}$. Each 2-way combination has four neuron-activation configurations $(0, 0)$, $(0, 1)$, $(1, 0)$ and $(1, 1)$. Among the six 2-way combinations, only the neuron-activation configurations of $\{n_1, n_2\}$, $\{n_1, n_4\}$, $\{n_2, n_3\}$ and $\{n_3, n_4\}$ are covered by $T$. Therefore, the 2-way combination sparse coverage for $T$ on $L_i$ is 66.6%.

**Table 1.** Activation status of $L_i$ on $T$

| $n_1$ | $n_2$ | $n_3$ | $n_4$ |
|---|---|---|---|
| 0 | 0 | 0 | 0 |
| 0 | 1 | 0 | 1 |
| 1 | 0 | 1 | 0 |
| 1 | 1 | 1 | 1 |

Since the $t$-way combination sparse coverage cannot take the coverage within each combination of neurons into account, next we will introduce the $t$-way combination dense coverage.

**Definition 3 ($t$-way combination dense coverage).** *For a set of test inputs $T$ and a set of neurons $L_i$, $t$-way combination dense coverage can be calculated as below.*

$$\text{TWdCov}(T, L_i) = \frac{\sum_{\theta \in \Theta} NC\_NA(T, \theta)}{2^t |\Theta|}$$

*Example 2.* Consider the neuron activation status of $L_i$ given test set $T$ in Table 1. Since there are six 2-way combinations of neurons in $L_i$ and each has four neuron-activation configurations, there are 24 neuron-activation configurations in total. The test set $T$ can cover 20 configurations and the neuron-activation configurations that are not covered are $\{n_1, n_3\} = (0,1)$, $\{n_1, n_3\} = (1,0)$, $\{n_2, n_4\} = (0,1)$ and $\{n_2, n_4\} = (1,0)$. Therefore, the 2-way combination dense coverage for $T$ is 83.3%. Recall that the 2-way combination sparse coverage for $T$ is only 66.6%.

**Definition 4 ($(p,t)$-completeness coverage).** *For a set of neurons $L_i$, $(p,t)$-completeness is defined as the probability that the $t$-ways combination dense coverage is at least $p$.*

*Example 3.* We again take Table 1 as an example. For 2-way combinations of neurons, the 2-way combination dense coverage for $\{n_1, n_2\}$, $\{n_1, n_4\}$, $\{n_2, n_3\}$ and $\{n_3, n_4\}$ are 100%, and 50% for $\{n_1, n_3\}$ and $\{n_2, n_4\}$. According to the Definition 4, we know that $(0.5, 2)$-completeness of $L_i$ is 100% and the $(1, 2)$-completeness for $L_i$ is 66.6%.

For simplicity, the combinatorial testing coverage defined above are for neurons within one layer. We can easily generalize those definitions in order to derive corresponding test coverage for neurons in multiple layers. The main idea is to perform the $t$-way combination of neurons within each layer. Then with all the formed neuron combinations $\Theta$, we will calculate different types of CT testing coverage.

## 4   CT Coverage Guided Robustness Testing of DL

Our proposed CT coverage criteria are general for both test suite evaluation and test generation guidance. In this study, we consider a typical problem of using CT criteria to test local adversarial robustness of DNNs designed for the purpose of classification [15,18]. Given an input **x** to a DNN, the local adversarial robustness property is concerned with whether there exists another input **x'** close enough to **x**, with respect to some distance metrics (e.g., $L_0$-norm, $L_\infty$-norm), such that **x** and **x'** are classified to different classes by the DNN. Such an input **x'**, once exists, is called an adversarial example of **x** and the DNN is not locally robust at **x**.

Let $\mathcal{C}(\mathbf{x})$ denote the class to which **x** is classified by DNNs. Formally, a DNN is $d$-locally-robust at an input **x** w.r.t a distance parameter $d$ iff we have the following [18]:

$$\forall \mathbf{x'} : ||\mathbf{x'} - \mathbf{x}|| \leq d \Rightarrow \mathcal{C}(\mathbf{x}) = \mathcal{C}(\mathbf{x'})$$

One approach to detecting adversarial example is through random testing. However, random testing often could be ineffective at detecting adversarial examples even if a large number of tests are generated [34]. To systematically generate tests to detect adversarial examples and analyze the local robustness of a given input, we propose to adapt combinatorial testing for test suite generation. Algorithm 1 shows the details of our CT coverage guided test generation. Given a seeded test set and $K$-way CT as input, the CT coverage table of the whole DNN is first initialized (see Line 1-2). Then the test generation iteration starts for each seeding test, guided by the CT coverage layer by layer. For each layer, the coverage is analyzed on the generated tests so far. The coverage table is updated and the uncovered CT targets of a layer $l$ are calculated (Line 6-7). After randomly selecting a target $ct_i$ to cover, we try to generate tests to cover $ct_i$ and analyze all the coverage targets they reach. Note that a test case can cover multiple CT targets in $K$-way testing, and the generated tests might not cover desired CT targets (Line 9-12). Before attempting on the next CT target, we check whether the generated tests contain adversarial examples and update the generated test suite for $T$, $T'$, and the test working set $T_{work}$ accordingly (Line 13). The test generation iteration continues until CT coverage targets are covered or processed, or the time limit hits.

In Algorithm 1, the test generation technique to cover the specific CT coverage target could be general without assumption on specific DNN internal structures (e.g., activation function types), such as search based testing [26], guided random testing [29], and symbolic execution and constraint-based testing [4]. To demonstrate the CT coverage guided test generation is helpful for detecting adversarial examples, this study assumes that DNN adopts the ReLU activation function and constraint solving based (i.e., by linear programming and CPLEX solver [16]) test generation [34]. In particular, a CT coverage target is encoded as the linear constraints with the object to minimize the $L_\infty$-norm perturbation distance on a seeding input test.

---

**Algorithm 1** Combinatorial-TestGen
**INPUT:** DNN $N$, seeding Test Set $T_s$, $K$-way
**OUTPUT:** TestSuite $T$, Adversarial Test Set $T'$

1: $T \leftarrow \{\}, T' \leftarrow \{\}$
2: $CT\_table \leftarrow initialize\_CT\_coverage\_table(T_s, K\text{-}way)$
3: **for** $t \in T_s$ **do**
4:     $T_{work} \leftarrow \{\}$
5:     **for** each layer $l \in N$ **do**
6:         $update\_CT\_coverage(CT\_table, T_{work})$
7:         $CT\_targets \leftarrow calculate\_CT\_targets(CT\_table, T_{work}, l)$
8:         **while** $CT\_targets \neq \emptyset$ **do**
9:             $target \leftarrow random\_select(CT\_targets)$
10:            $gen\_tests \leftarrow TestGen(t, target)$
11:            $covered\_targs \leftarrow cal\_cov\_targets(CT\_table, T_{work}, gen\_tests)$
12:            $CT\_targets \leftarrow CT\_targets - (covered\_targs \cup target)$
13:            $update\_TestSuite(gen\_tests, T_{work}, T, T')$
14: **return** $T, T'$

---

## 5   Evaluation

We have implemented *DeepCT*, a DL combinatorial testing framework that performs automated test generation for DNNs based on Keras (ver.2.1.3) [8] and Tensorflow (ver.1.5.0) [1]. In the current version, *DeepCT* provides a LP constraint solving based test generation, and we adopt it to investigate whether CT and our proposed criteria are useful for testing DNNs.

In this study, we mainly investigate whether *DeepCT* and our criteria are useful to guide testing towards the robustness detection of DNNs. We use the publicly available dataset MNIST [21] and two pre-trained DNN models. In particular, the two studied DNNs contain 3 (64*32*64 with 55,082 parameters) and 5 (84*42*64*42*84 with 79,454) fully-connected hidden layers, and obtain 99.965%, 99.872% training accuracy, and 97.63%, 97.51% test accuracy respectively. For the DNNs' local robustness analysis, we randomly seed 1,000 tests which can be correctly handled by our studied DNNs from MNIST accompanied test sets as the study subject.

All experiments were run on a high performance computer cluster. Each cluster node runs a GNU/Linux system with Linux kernel 3.10.0 on a 18-core 2.3GHz Xeon 64-bit CPU with 196 GB of RAM.

### 5.1  Random Testing

Although previous work [34] advocated that random testing is ineffective in detecting the local robustness issues of DNNs, we believe random testing is easy to use and scalable, which is always worth a first shot before further in-depth analysis. Therefore, our first step performs random test generation to analyze the robustness of two studied DNNs (i.e., $DNN_1$ and $DNN_2$) on the 1,000 seeded tests. To be specific, we randomly generate 10,000 tests for each seeded test and analyze whether robustness issues could be detected.[6] The experiment results show that the random testing is already able to detect robustness issues on 194 seeded tests on $DNN_1$ and 178 on $DNN_2$ with a total of 266 unique issues, 106 of which are shared issues on both $DNN_1$ and $DNN_2$. This is consistent with our intuition that the robustness of a DNN on handling different test input could be different. A test input near a DNN's decision boundary could cause robustness issues more easily. Our experimental results confirm this observation and indicate that random testing could already be useful to detect robustness issues on some fragile test input.

> *Random Testing* can be useful to detect local robustness issues and worth a first shot before further in-depth analysis.

### 5.2  DeepCT

For the obtained 1,000 seeded tests in Section 5.1, we first filter out those 266 tests whose robustness issues can already be detected by random testing. We found that if a test input $t$ could be detected as locally-non-robust, $t$ is often quite fragile to random perturbation. Random testing would often find quite a number of adversarial examples for $t$. Inadvertently including such adversarial examples into statistics in line with other test inputs (those random testing could not detect robustness issues) would pollute the overall analysis results, causing the overall adversarial example ratio of the generated tests to be seemingly high.

---

[6] Each of 784 pixels of the test image is normalized to range $[0, 1]$, and randomly perturb each pixel within range $[-0.15, 0.15]$, that is 0.15-locally-robustness analysis.

**Table 2.** The obtained CT coverage and detected adversarial examples for random testing, and CT testing.

|  | Testing Method |  | Combinatorial Testing Coverage (%) | | | | #Accu. Tests | Adv. Ratio(%) |
|---|---|---|---|---|---|---|---|---|
|  |  |  | 2-Way Spar. | 2-Way Den. | (0.5,2)-C. | (0.75,2)-C. | | |
| $DNN_1$ | Random | | 2.28 | 34.95 | 33.75 | 3.75 | 10,000 | 0.00 |
|  | CT | $L_1$ | 60.27 | 81.56 | 95.01 | 70.98 | 4,073 | 0.29 |
|  | CT | $L_2$ | 76.94 | 91.98 | 99.67 | 91.30 | 6,768 | 2.17 |
|  | CT | $L_3$ | 93.62 | 98.23 | 100.00 | 99.32 | 8,032 | 9.91 |
| $DNN_2$ | Random | | 1.18 | 32.56 | 26.98 | 2.10 | 10,000 | 0.00 |
|  | CT | $L_1$ | 46.96 | 75.10 | 91.95 | 61.50 | 8,547 | 1.87 |
|  | CT | $L_2$ | 68.91 | 87.52 | 98.64 | 82.55 | 11,573 | 3.53 |
|  | CT | $L_3$ | 97.15 | 99.05 | 100.0 | 99.03 | 13,129 | 8.84 |
|  | CT | $L_4$ | 97.41 | 99.11 | 100.0 | 99.03 | 13,217 | 9.35 |
|  | CT | $L_5$ | 97.81 | 99.21 | 100.0 | 99.03 | 13,351 | 9.98 |

For the remaining 734 tests, we randomly sample 50 tests for further in-depth analysis on tests generated by *DeepCT* to analyze the $d$-locally-robustness (where $d = 0.15$, and $t = 2$ for neuron combinations, also see Algorithm 1) in line with the tests generated by random testing. Table 2 summarizes the achieved averaged coverage results and accumulated generated tests. For each corresponding test generation method, Columns 4-7 show the obtained coverage of 2-way combination sparse, 2-way dense, (0.5,2)-completeness, and (0.75,2)-completeness coverage. For combinatorial testing, *DeepCT* incrementally generates tests to cover CT coverage targets layer by layer, Column 8 gives the accumulated number of generated tests and Column 9 shows the corresponding detected adversarial ratio of the generated tests.

Overall, for both studied DNNs, random testing achieves fairly low coverage of all the evaluated coverage, and *DeepCT* achieves rather high coverage as more layers are tested. In addition, we see that, for *DeepCT*, all of the 50 studied tests are not 0.15-locally-robust for both DNNs, and such issues could not be detected by random testing quite a number of (i.e., 10,000) tests are generated. Table 2 shows that random testing achieves only 2.28% and 1.18% 2-way sparse coverage on two studied DNNs. Compared with (0.5,2)-completeness coverage, (0.75,2)-completeness coverage is also much lower, which indicates random testing does not deeply cover many of the neuron activation configurations of 2-way neuron combinations. This might be because that random testing lacks some discipline to systematically explore the states of DNN, where the adversarial cases lie in. Therefore, in cases where random testing could not detect the local-robustness issues, it is still unable to confirm the robustness on an input with some confidence.

> *Random Testing* does not provide confidence when local-robustness cannot be detected.

In comparison with random testing, *DeepCT* obtains 60.27% and 46.96% 2-way sparse coverage even only the first hidden layer $L_1$ is tested and analyzed, with a reasonable number of test size while already being able to detecting adversarial examples for all seeded tests. In our evaluation, 4,073 and 8,547 tests (on average) are generated for $DNN_1$ and $DNN_2$, respectively. This is a rather small number of tests compared with all the possible neuron combination runtime states, but still enables to detect sufficient adversarial examples. As described in Algorithm 1, when *DeepCT* analyzes the second hidden layer $L_2$, it first analyzes the coverage obtained by tests generated by all previous layers, and generates tests to only cover the remaining uncovered targets on layer $L_2$. In the two previously studied DNNs, we find that after generating tests for layer $L_2$, 2-way sparse combination coverage increases by 16.67% and 21.95%, respectively. Similar coverage boost could also be observed by other coverage criteria. For example, the obtained (0.75,2)-completeness coverage (i.e., 91.30% and 99.03%) indicates that the 2-way neuron interactions are mostly covered deeply for both DNNs. When the first three layers of both DNNs are analyzed, the 2-way sparse coverage reaches 93.62% and 97.15%, respectively, where most of the adversarial examples might already be detected. For $DNN_2$, only about 220 new tests in total are created when analyzing layers $L_4$ and $L_5$. As for the detected adversarial examples, adversarial ratio gaps of $L_2$ and $L_3$ are much larger than other layers for both DNNs. This might indicate that the different layers might contribute differently to detect adversarial examples by CT, and some layers should be intensively covered than others. In our studied DNNs, CT-guided test generation that systematically covers the CT coverage targets for the first several layers of DNNs already enables to detect sufficient adversarial examples and local-robustness issues; we will investigate whether this results could generalize to other dataset and DNNs in our future work.

> Our proposed combinatorial testing coverage criteria are useful for adversarial example detection and local-robustness analysis.
> *DeepCT* enables the guided test generation to achieve high CT coverage and allows to detect local-robustness issues even only the first several layers of a DNN are analyzed.

### 5.3  Discussion

In this paper, we demonstrate the usefulness of combinatorial testing for DNNs, and we believe CT would be an important and promising direction for testing DNNs. However, many open questions need further investigation towards the practical application of CT to real-world large-scale DL systems. In the empirical study of CT for traditional software, Kuhn et.al. [19] found that testing the interactions of a few parameters already enables to find many software defects.

In our study, we evaluated the 2-*way* CT cases for DNNs, and confirms that similar conclusions might also be applied to testing DNNs. One of our future work would focus on the understanding and interpretation of how the parameter $t$ influences the $t$-*way* defect detection ability, and the corresponding confidence for the analysis of the local-robustness of DNNs.

As mentioned in the last section, *DeepCT* proposes a general combinatorial coverage guided testing method for DNNs, the test generation technique could be substituted by other techniques as well. The purpose of this work that adopts a constraint-based approach is to demonstrate the usefulness of the proposed CT coverage criteria for testing DNNs. However, we find that the number of neurons is often quite large in practical DL systems, and it would greatly hinder the constraint-based test generation techniques. Although solvers like CPLEX [16] already represents the state-of-the-practice, its scalability to real-world DLs is still a big concern. We will perform further in-depth study on more efficient and scalable test generation techniques in our future work.

In this study, we provide one typical strategy to generate combinatorial covering targets to guide the CT test generation as the first attempt. Many advanced combinatorial test generation strategies could be further explored, which could potentially obtain high CT coverage with even less tests [10, 13, 17, 22, 27, 40]. We will also study diverse more optimized CT strategies in the next step.

## 6  Conclusion

Combinatorial testing is a well-established and successful technique in traditional software testing. Rather than exhaustively searching all the combinations of input space, CT focuses on testing the interactions of inputs, aiming to reduce the test size while obtaining satisfiable defect detection abilities. This paper initiates an empirical study on the usefulness of CT for testing DL systems. Our evaluation results demosrate that CT provides a promising avenue for testing DL systems.

We again emphasize that our research is the first to adapt the concept of CT, provide CT coverage guided test generation, and perform an empirical study toward evaluating the robustness of DL systems using CT. We thus hope that applying CT to DL systems may constitute a seminal foresight built on our evaluation results, and in parallel shed light on the construction of more CT coverage criteria and scalable test generation techniques towards achieving robust real-world DL applications.

## Acknowledgement

This research is partially supported by a recently awarded grant *Robust Deep Learning and its Application to high Confidence Medical Diagnosis*, which is a pivotal sub-project of *Chinese 100K Human Genome Project* (i.e., a National Key R&D Program of China, No. 2017YFC1201100). The overall goal of this sub-project includes both (1) Fundamental research: design and explore diverse

12      L. Ma et al.

potentially useful testing, verification methods (including metric criteria) for DL systems, and further propose solutions to construct robust DL systems; and (2) Applied research: design typical robust DL system for fully automated, high confidence, and large-scale biomedical classification, and medical diagnosis.